\documentclass[3p,twocolumn,sort&compress]{elsarticle}

\usepackage{lineno,amsmath,amsfonts} 
\usepackage{tikz}
\usetikzlibrary{plotmarks}
\usetikzlibrary{arrows}
\usetikzlibrary[patterns]
\usetikzlibrary{shapes.geometric,arrows}
\modulolinenumbers[5]

\usepackage[final]{changes} 
\definechangesauthor[color=blue]{NB} 
\definechangesauthor[color=red]{FZ}

\usepackage{soul}
\soulregister\cite7
\soulregister\ref7
\soulregister\eqref7

\usepackage{hyperref}
\hypersetup{
    colorlinks=true,
    linkcolor=blue,
    filecolor=magenta,      
    urlcolor=cyan,
}

\newcommand{\vect}[1]{\boldsymbol{#1}}
\newcommand{\tensor}[1]{\boldsymbol{#1}}

\journal{arXiv}

\begin{document}

\begin{frontmatter}

\title{Learning dislocation dynamics mobility laws from large-scale MD simulations}

\author[llnl]{Nicolas Bertin\corref{cor1}}
\ead{bertin1@llnl.gov}

\author[llnl]{Vasily V. Bulatov}
\author[llnl]{Fei Zhou}

\cortext[cor1]{Corresponding author}
\address[llnl]{Lawrence Livermore National Laboratory, Livermore, CA, USA}

\begin{abstract}

The computational method of discrete dislocation dynamics (DDD), used as a coarse-grained model of true atomistic dynamics of lattice dislocations, has become of powerful tool to study metal plasticity arising from the collective behavior of dislocations.
As a mesoscale approach, motion of dislocations in the DDD model is prescribed via the mobility law; a function which specifies how dislocation lines should respond to the driving force.
However, the development of traditional ``hand-crafted'' mobility laws can be a cumbersome task and may involve detrimental simplifications.
Here we introduce a machine-learning (ML) framework to streamline the development of data-driven mobility laws which are modeled as graph neural networks (GNN) trained on large-scale Molecular Dynamics (MD) simulations of crystal plasticity.
We illustrate our approach on BCC tungsten and demonstrate that our GNN mobility implemented in large-scale DDD simulations accurately reproduces the challenging tension/compression asymmetry observed in ground-truth MD simulations \added{while correctly predicting the flow stress at lower straining rate conditions unseen during training}, thereby demonstrating the ability of our method to learn relevant dislocation physics.
Our \added{DDD+ML} approach opens new promising avenues to improve fidelity of the DDD model and to incorporate more complex dislocation motion behaviors in an automated way, \added{providing a faithful proxy for dislocation dynamics several orders of magnitude faster than ground-truth MD simulations}.

\end{abstract}

\end{frontmatter}


\section{Introduction}

In metals, plastic deformation is ordinarily defined by the motion and interaction of dislocation lines through the lattice. While direct Molecular Dynamics (MD) simulations of crystal plasticity are now within reach to provide insights on the collective behavior of dislocations and metal strength \cite{zepeda2017probing, zepeda2021atomistic, bertin2022sweep, stimac2022energy}, their scale still commands considerable computing resources.
Alternatively, the  method of Discrete Dislocation Dynamics (DDD), used as an expedient mesoscale proxy, has been widely regarded as a powerful tool to connect the evolution of the dislocation microstructure to the macroscopic response of crystals, thereby bridging the gap between mesoscale and continuum approaches in metal plasticity \cite{kubin1992dislocation, zbib1998plastic, bulatov1998connecting, schwarz1999simulation, ghoniem2000parametric, weygand2002aspects, Arsenlis07}. In contrast to MD which simulates ``all the atoms'', only the dislocation microstructure is considered in DDD, which is represented as a set of dislocation segments inter-connected through nodes and generally evolved with an over-damped equation of motion \cite{BulatovCai06}
\begin{equation} \label{eq:eom}
    \vect{V}_i = \frac{d\vect{r}_i}{dt} = \mathcal{M}\left[ \vect{F}_i\left( \{\vect{r}_j\}, \{ \vect{b}_{jk} \} \right) \right]
\end{equation}
where $\vect{r}_j \in \mathcal{V}$ and $\vect{b}_{jk} \in \mathcal{E}$ are the dislocation node positions and Burgers vector connectivity, respectively defining the set of vertices and edges of the  dislocation graph $\mathcal{G} = (\mathcal{V}, \mathcal{E})$.
In Eq.~\eqref{eq:eom}, $\mathcal{M}$ is the mobility function that relates the nodal force $\vect{F}_i$ exerted on the dislocation line at node $i$ to its resulting nodal velocity $\vect{V}_i$. The mobility law is the core ingredient of the DDD model which encapsulates the physics of the simulation (analogous to the interatomic potential in MD), i.e. by which DDD practitioners specify how dislocations should move in response to the driving force.

In principle, the DDD mobility needs to be calibrated to a reference behavior in order for simulations to produce meaningful predictions. Given the relative difficulty to extract force-velocity relations from experiments, a widely used strategy has been to calibrate mobility laws using MD simulations of individual dislocations.
Within this approach, a mobility law is typically constructed by postulating a ``hand-crafted'' functional form based on physical knowledge and intuition, and quantifying its parameters by running MD calculations of isolated dislocations \cite{wang2011atomistically, srivastava2013dislocation, po2016phenomenological}, e.g. by computing the velocity of dislocations of distinct types (for instance, edge and screw) under varying levels of stresses and temperatures \cite{chang2001dislocation, olmsted2005atomistic, queyreau2011edge, cereceda2013assessment, cho2017mobility}, extracting nucleation energy barriers and critical stresses for motion controlled by the kink-pair mechanism \cite{monnet2009structure, kang2012singular, po2016phenomenological}, or extracting dislocation drag by analysis of the thermal fluctuations spectrum \cite{geslin2018thermal}.

However, while this conventional workflow has been widely adopted, it presents several important drawbacks.
First, constructing mobility laws can be a time-consuming task, and differences in metals crystallography and dislocation behaviors make it difficult to fully automate the process.
Second, this approach necessarily involves simplifications. In particular, it remains unclear if and how behaviors extracted from straight, isolated dislocations in MD can be transferred \cite{bertin2020frontiers} to large-scale DDD simulations in which dislocations form dense and intricate networks of curved lines \cite{sills2018dislocation}.
Third, to the authors' knowledge, no work has yet verified how DDD predictions using mobility laws constructed in the way described above would compare to direct, large-scale MD simulations performed using the same interatomic potential used in the calibration.

In this work, we introduce \added{the DDD+ML} framework to streamline the construction of DDD mobility laws using a machine learning (ML) approach akin to the task of developing interatomic potentials for MD simulations.
For this, we propose to define the mobility law in Eq.~\eqref{eq:eom} as a graph neural network (GNN) operator trained to learn the evolution of dislocation network trajectories produced by large-scale MD simulations.
Here, we therefore view DDD as a coarse-grained model of direct MD simulations of crystal plasticity, which we regard as the ground-truth that we wish to emulate.
Thus, a faithful mobility law should result in \added{DDD+ML} predictions that closely match those of direct MD simulations performed under identical conditions.
Using body-centered cubic (BCC) W as a testbed material, we show that our trained GNN-based mobility implemented in our DDD model can learn relevant dislocation physics and is able to reproduce well the strength behavior and tension/compression asymmetry predicted in large-scale MD simulations.
We believe that our new GNN-based mobility framework provides a novel, unbiased approach to accurately incorporate dislocation physics in the DDD model while paving the way to automated workflows for constructing mobility laws of arbitrary complexity.

\section{Results} \label{sec:results}

\subsection{\added{DDD+ML} model} \label{sec:GNNmodel}

\begin{figure}[t]
  \begin{center}
    \includegraphics[width=0.48\textwidth]{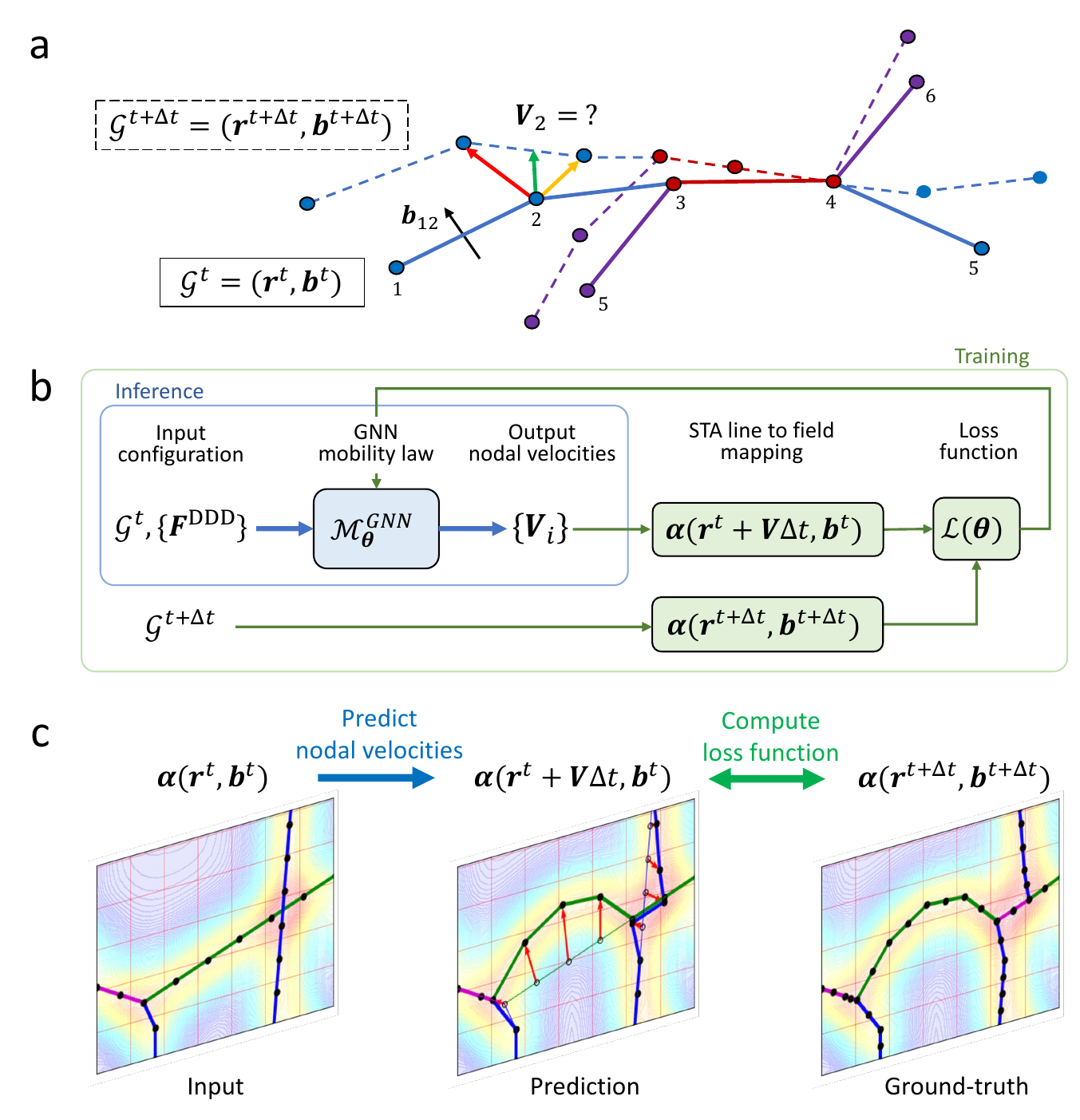}
  \end{center}
  \vspace*{-0.3cm}
  \caption{DDD+ML framework. (a) Schematic of two consecutive dislocation networks $\mathcal{G}^t$ and $\mathcal{G}^{t+\Delta t}$ extracted from a MD trajectory using DXA at times $t$ (solid lines) and $t+\Delta t$ (dashed lines). As extracted, the configurations contain nodal positions $\vect{r} = \{\vect{r}_i\}$ and Burgers vector connectivity $\vect{b} = \{\vect{b}_{ij}\}$ but (i) contain no information about the line forces driving their evolution, and (ii) nodal velocity vectors generally cannot be unambiguously defined due to the nonisomorphism between both networks. For instance, colored arrows show three possible ways to define the ground-truth velocity $\vect{V}_2$ at node $2$. (b) Schematic of the inference and training loop for the GNN-based mobility law. 
\added{(c) Illustration of the Nye's tensor field-matching approach to circumvent the ill-defined matching of velocities. As exemplified here, the procedure is agnostic to line discretization and network topology.}}
\label{fig:model}
\end{figure}

Taking advantage of the graph nature of dislocation networks, we view the construction of the DDD mobility law in Eq.~\eqref{eq:eom} as a ML task defined on a graph network, Fig.~\ref{fig:model}.
\added{Very recently, we successfully developed a machine-learned mobility from a simplified dislocation-obstacle DDD system \cite{bertin2023accelerating}. Here, we propose to model} the mobility function with a GNN trained to predict the evolution of subsequent dislocation network configurations, e.g. as extracted from large-scale (ground-truth) MD trajectories using the dislocation extraction algorithm (DXA) \cite{Stukowski10, Stukowski14}.

In practice however, the task is complex and presents several challenges that need to be addressed. To illustrate the matter, consider two consecutive dislocation graphs $\mathcal{G}^t$ and $\mathcal{G}^{t+\Delta t}$ at time $t$ and $t+\Delta t$, respectively, as depicted in Fig.~\ref{fig:model}a. As implied by Eq.~\eqref{eq:eom}, training a ML mobility law entails learning the relation between the (input) nodal force and (output) nodal velocity vectors given the geometry and topology of the dislocation network. Yet, by themselves dislocation networks extracted with DXA (i) contain no information about the driving force on the dislocation lines, and (ii) the dislocation velocities are only implicitly defined by comparing dislocation line configurations attained at consecutive times, and thus not explicitly contained in the data. This is because the topology of dislocation networks constantly evolves (e.g. as a result of dislocation intersections and core reactions) while the discretization of dislocation links is essentially arbitrary, i.e. subsequent dislocation graphs are generally nonisomorphic.
In other words, there does not exist a one-to-one correspondence between dislocation nodes in consecutive network configurations \added{and} velocity vectors cannot be unambiguously defined. As such, learning on nodal velocity vectors is not a well-defined task for ML.

We propose to address both these challenges of unknown ground-truth force and velocity vectors in the following way. First, we estimate the nodal forces $\vect{F}_i$ on nodes $i$ of the MD dislocation networks using the force calculation procedure of the DDD model
\begin{equation} \label{eq:Fi_DDD}
    \vect{F}_i \sim \vect{F}_i^{\rm DDD} = \vect{F}_i^{\rm app} + \vect{F}_i^{\rm lr} + \vect{F}_i^{\rm sr} + \vect{F}_i^{\rm core}
\end{equation}
where the force $\vect{F}_i^{\rm DDD}$ calculated with DDD is the sum of several contributions, namely the force due to the applied loading and boundary conditions $\vect{F}_i^{\rm app}$, the long-range $\vect{F}_i^{\rm lr}$ and short-range $\vect{F}_i^{\rm sr}$ elastic interactions, and the force associated with the dislocation core $\vect{F}_i^{\rm core}$.
Given the various approximations involved in force calculations in the DDD model, this force is only an \emph{estimate} of the true (ground-truth) but unknown force $\vect{F}_i^{\rm MD}$ driving the dislocations in the MD simulations. Mathematically, we can write this approximation exactly as
\begin{equation} \label{eq:Fi_MD}
    \vect{F}_i^{\rm MD} = \vect{F}_i^{\rm DDD} + \vect{F}_i^{\rm corr}
\end{equation}
where $\vect{F}_i^{\rm corr}$ is a correction term accounting for the error made in approximating the true force with the DDD model.
Assuming that the applied and long-range components of the force are smoothly varying functions that can be accurately computed with DDD, it follows that the correction term in Eq.~\eqref{eq:Fi_MD} is a rather local function of dislocation network, which can itself be easily learned by a GNN model.
Thus, we propose to define a GNN-based mobility law $\mathcal{M}^{\rm GNN}_{\vect{\theta}}$ with learnable parameters $\vect{\theta}$,
\begin{equation} \label{eq:M_GNN}
    \vect{V}_i = \mathcal{M}^{\rm GNN}_{\vect{\theta}}\left( \vect{F}_i^{\rm DDD}, \{ N_j \}, \{ E_{jk} \} \right),
\end{equation}
and let the model learn the resulting nodal velocity vector $\vect{V}_i$ as a function of the estimated force $\vect{F}_i^{\rm DDD}$ and local graph neighborhood of nodes $j$ and segments $jk$ with attributes $N_j$ and $E_{jk}$, respectively (see Methods).
As defined, the mobility function in Eq.~\eqref{eq:M_GNN} thus fully subsumes the calculation of the as-yet unspecified local force correction. Since in DDD simulations forces are always computed using Eq.~\eqref{eq:Fi_DDD}, this formulation further ensures that the approach is self-consistent when employing the trained mobility in rollout simulations.

We now turn our attention to the second issue, namely the absence of well-defined nodal velocity vectors in the ground-truth MD data. Here we note that this issue is essentially identical to the task of tracing dislocation motion between consecutive MD dislocation snapshots, for which we have recently introduced the sweep-tracing algorithm (STA) \cite{bertin2022sweep}. By using a dual line/field representation of the dislocation networks, STA can ``reconnect'' successive networks in a way that is independent of line discretization and agnostic to network topology. Conceptually, the task is defined as an optimization problem that seeks to minimize the distance between Nye's tensor \cite{Nye53} field representations of consecutive graphs \cite{bertin2022sweep}.
Following the STA approach, the training of the GNN-based mobility operator in Eq.~\eqref{eq:M_GNN} is achieved by minimizing the following loss function
\begin{multline} \label{eq:loss}
    \mathcal{L}(\vect{\theta}) = \sum_s \sum_{\vect{g}} \sum_{kl} \Bigl[  \alpha^{\vect{g}}_{kl}\left( \vect{r}^{t_s} + \vect{\mathcal{M}}_{\vect{\theta}}^{\rm GNN} \Delta t_s, \vect{b}^{t_s} \right) \\ - \alpha^{\vect{g}}_{kl}\left( \vect{r}^{t_s+\Delta t_s}, \vect{b}^{t_s+\Delta t_s} \right) \Bigr]^2
\end{multline}
over the set of training examples $s$, where $\alpha_{kl}^{\vect{g}}$ are the components of the Nye's tensor field numerically computed at grid point $\vect{g}$ using the method introduced in \cite{bertin2019connecting}.
During training, the mobility law thus learns to predict nodal velocity vectors $\vect{V}_i$ which, when applied to configurations at time $t_s$, best match the tensor field representation of the ground-truth evolution of the networks at time $t_s+\Delta t_s$, Fig.~\ref{fig:model}c. In other words, the field-matching loss of Eq.~\eqref{eq:loss} circumvents of the seemingly straightforward but ill-defined matching of velocities.

We point out that the calculation of the Nye's tensor field over networks is only required to compute and back-propagate mobility parameters $\vect{\theta}$ information during training. Once trained, only the GNN-based mobility function in Eq.~\eqref{eq:M_GNN} needs to be evaluated at inference time, Fig.~\ref{fig:model}b.
Our GNN model and training protocol are implemented using PyTorch \cite{pytorch}.

\subsection{Validation on DDD trajectories}

To validate our approach, we first apply our method to DDD trajectories to evaluate the ability of our framework to learn a simple ``hand-crafted'' DDD mobility law. Although the approach was described for DXA configurations extracted from MD in \S\ref{sec:GNNmodel}, it can be applied equally well to DDD configurations. Doing so can be insightful since, in contrast to MD simulations, we have complete control over traditional mobility laws in DDD and can prescribe any desired ground-truth force-velocity relation, and then examine how well our approach can recover it. Here we choose a generic linear mobility function developed for BCC metals \cite{cai2004mobility} and implemented in our DDD code ParaDiS \cite{Arsenlis07}, which can be expressed as
\begin{equation}
    \vect{V}_i = \tensor{\mathcal{M}}^{\rm DDD}(M_s, M_e) \cdot \vect{F}_i^{\rm DDD}
\end{equation}
where $\tensor{\mathcal{M}}^{\rm DDD}$ is a matrix parameterized by scalar mobility coefficients $M_s = 20~({\rm Pa}\cdot{\rm s})^{-1}$ and $M_e = 2600~({\rm Pa}\cdot{\rm s})^{-1}$ associated with the velocity of pure screw and pure edge dislocations, respectively.

To generate the training data, we run 5 large-scale DDD simulations each starting with 12 prismatic loops seeded at different random positions. Crystals are compressed at a strain rate of $2\times 10^8$/s along the [001] direction. Simulations are run to 0.1 strain during which the dislocation networks are regularly saved at intervals of 0.25~ps, generating a total of 10,000 configurations for training. These configurations then follow the same procedure as the one described for DXA snapshots in \S\ref{sec:GNNmodel}: first nodal forces are computed using DDD, and then training is performed by minimizing the STA-based loss function in Eq.~\eqref{eq:loss}.

Once trained, we inspect the best resulting GNN-based mobility law by inferring the nodal velocities of pure edge and screw dislocations as a function of shear stress. For this, we create small infinite, straight dislocation configurations of edge and screw characters and compute the nodal forces $\vect{F}_i^{\rm DDD} = \vect{F}_i^{\rm app}(\tau)$, where $\tau$ is the probing shear stress value. This simple technique allows us to slice through the otherwise ``black-box'' nature of the model.

\begin{figure}[t]
  \begin{center}
    \includegraphics[width=0.45\textwidth]{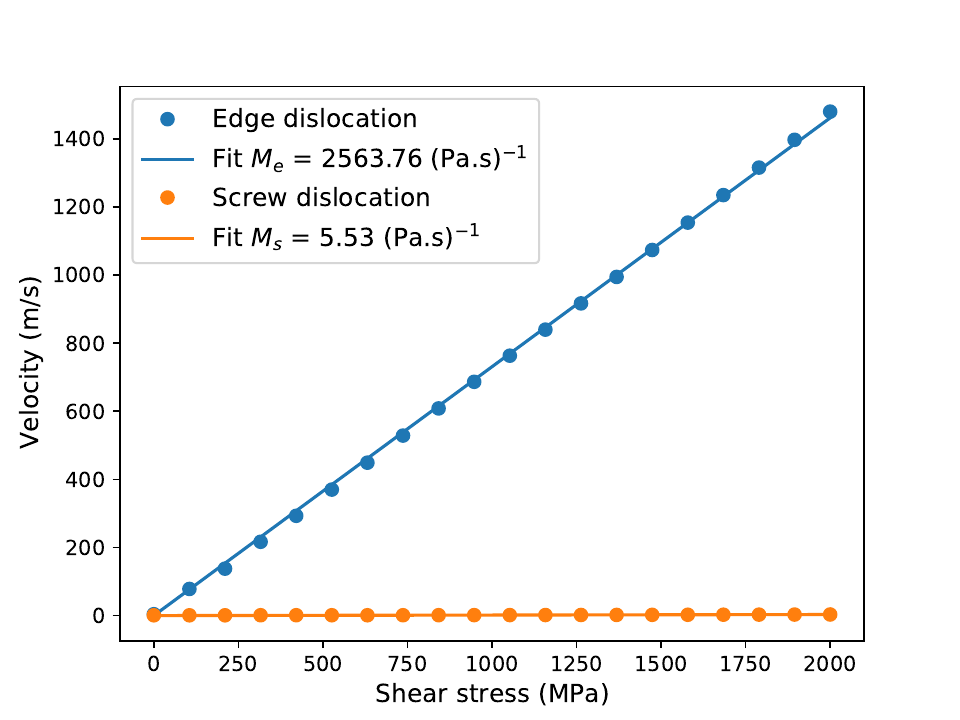}
  \end{center}
  \vspace*{-0.5cm}
  \caption{Edge and screw dislocation velocities predicted by the GNN-based mobility law trained on DDD trajectories. The GNN model does capture the linear nature of the ground-truth DDD mobility law, with linear fits of the stress-velocity curves that give mobility coefficients in good agreement with ground-truth values.}
\label{fig:ddd_mob}
\end{figure}

Results in Fig.~\ref{fig:ddd_mob} show the velocity magnitudes as function of shear stress predicted by the trained GNN-mobility law. The linear nature of the ground-truth DDD mobility law is well captured by the GNN model. In addition, fitting the stress-velocity curves with a linear regression yields mobility coefficients of $\hat{M}_e \approx 2564~({\rm Pa}\cdot{\rm s})^{-1}$ and $\hat{M}_s \approx 6~({\rm Pa}\cdot{\rm s})^{-1}$ for the edge and screw dislocations, respectively, which are in good agreement with the ground-truth values.
While further training could lead to yet better accuracy, this simple example demonstrates the general feasibility of our proposed framework.

\subsection{Application to MD trajectories} \label{sec:md_train}

We now apply our method to learn a mobility law from large-scale MD simulations. To illustrate the approach, we focus on the case of BCC plasticity which is notoriously complex due to the strong plastic anisotropy \cite{christian1983some, duesbery1998plastic, dezerald2016plastic}, as manifested in the tension/compression asymmetry \cite{sherwood1967plastic, webb1974effect, bertin2022sweep}, and the still debated issue of slip crystallography \cite{weinberger2013slip}.
We choose BCC W as a testbed material because it is nearly isotropic, which makes it convenient for the use of isotropic elasticity formulations in the DDD calculations. MD simulations are run using the EAM-style interatomic potential developed in \cite{Juslin13}, which yields an anisotropic ratio $A = 2C_{44}/(C_{11}-C_{12}) \sim 1.15$ at 300~K.

We perform 2 large-scale simulations of $\sim 35$ millions atoms in which BCC crystals initially seeded with prismatic dislocation loops are deformed at 300~K under uniaxial tension and compression along the [001] direction. Crystals are deformed at a true strain rate of $2\times 10^8$/s until reaching a strain of 1.0. Consistent with our previous results in BCC Ta \cite{bertin2022sweep, bertin2023crystal}, the simulations predict a strong tension/compression asymmetry, with flow stress of $\sim 4.2$~GPa in compression and $\sim 2.8$~GPa in tension.
During both runs, DXA is used to extract the dislocation configurations at every 1~ps interval of time. This results in a total of 10,000 DXA snapshots which are then used to train a GNN-based mobility.

\begin{figure}[t]
  \begin{center}
    \includegraphics[width=0.45\textwidth]{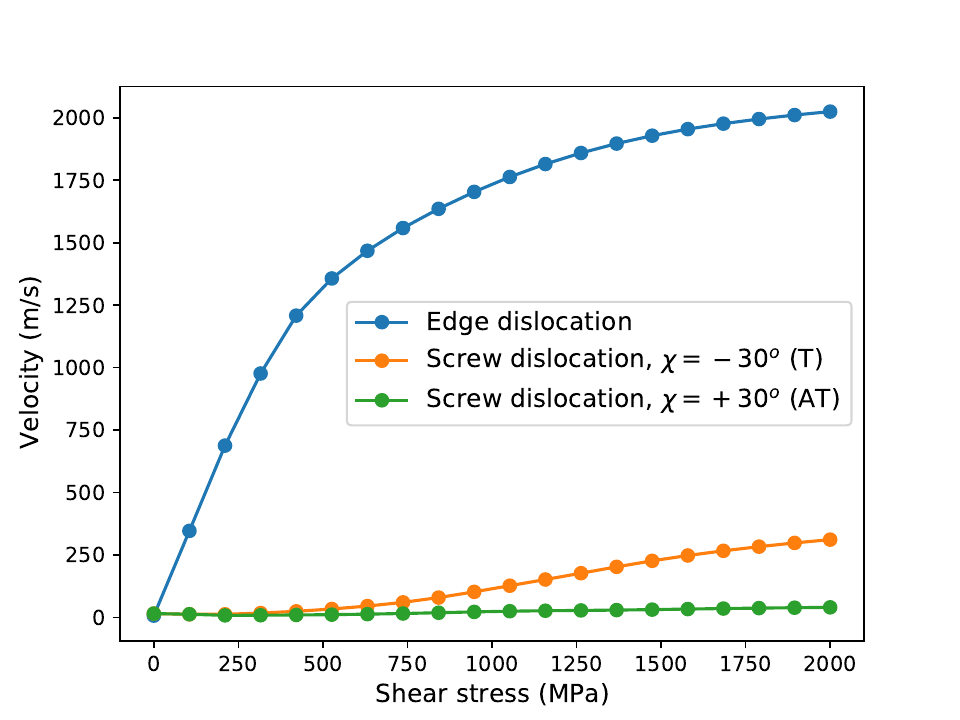}
  \end{center}
  \vspace*{-0.5cm}
  \caption{Edge and screw dislocation velocities predicted by the GNN-based mobility law trained on large-scale MD trajectories of BCC W deformed under [001] tension and compression. The GNN mobility law predicts a non-linear force-velocity relation for the edge dislocation and a strong asymmetry in the velocity of the screw dislocation sheared in the twinning (T) and anti-twinning (AT) direction.}
\label{fig:md_mob}
\end{figure}

Results for edge and screw dislocation velocities as a function of shear stress predicted by the best trained GNN-based mobility law are shown in Fig.~\ref{fig:md_mob}. The model exhibits smooth and well-behaved functions. The predicted edge dislocation velocity shows a linear stress-velocity relation up to $\sim 300$ MPa, followed by a non-linear regime reaching an asymptotic velocity of $\sim 2000$~m/s at a stress of 2~GPa. Qualitatively, this result is fully consistent with the known behavior of the edge dislocation \cite{po2016phenomenological} and previous MD calculations in BCC metals \cite{chang2001dislocation, osetsky2003atomic}.

Of particular interest are the predicted velocities for the screw dislocation. When sheared along the twinning (T) direction (corresponding to a MRSSP angle $\chi = -30^o$), the screw velocity response resembles that of thermally-activated motion, with very slow velocity at stresses below an activation threshold (finite temperature Peierls stress), and a transition to a drag-controlled regime at higher stresses. In contrast, when sheared in the anti-twinning (AT) direction (corresponding to a MRSSP angle $\chi = +30^o$), the screw dislocation is predicted to be much slower on average. Here again these results are fully consistent with theoretical models \cite{edagawa1997motion} and earlier observations \cite{bertin2022sweep}.

\subsection{Large-scale \added{DDD+ML} simulations with trained mobility} \label{sec:ddd_md_results}

To assess the validity of our approach and the quality of the learned model in realistic conditions, we now examine predictions of large-scale \added{DDD+ML} simulations using the GNN-based mobility law trained in \S\ref{sec:md_train}, which is implemented within the ParaDiS DDD code \cite{Arsenlis07} via the C++ PyTorch interface.
To establish a one-to-one comparison, DDD simulations are performed under identical conditions to those used in the ground-truth MD simulations.
Cubic DDD simulation boxes of side length $296b$, where $b=0.2743$~nm is the magnitude of the Burgers vector, are initially seeded with randomly positioned prismatic loops (using different seeds than in the ground-truth MD runs) and then deformed under [001] tension and compression at a strain rate of $2\times 10^8$/s. We use values of the shear modulus $\mu = 149.78$~GPa and Poisson's ratio $\nu = 0.289$, corresponding to values of the elastic constants of the ground-truth interatomic potential \cite{Juslin13} at 300~K. Dislocation core forces are computed from core energies extracted from the same interatomic potential using the framework detailed in \cite{bertin2021core}.

\begin{figure}[t]
  \begin{center}
    \includegraphics[width=0.49\textwidth]{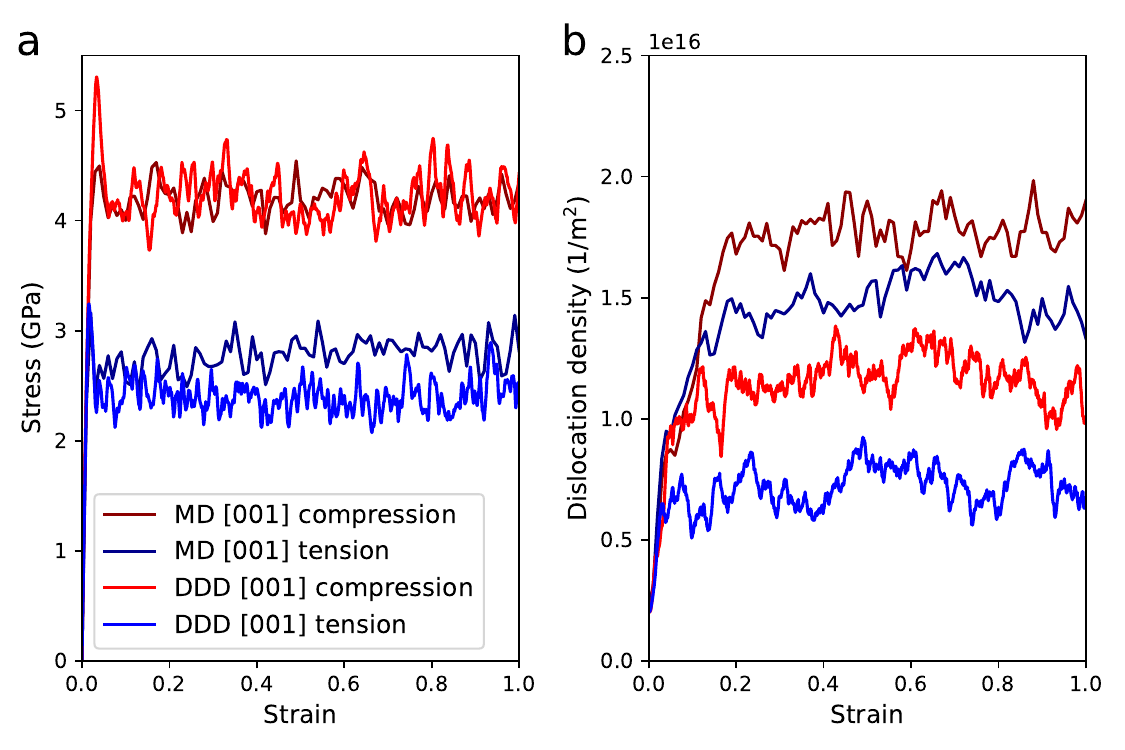}
  \end{center}
  \vspace*{-0.5cm}
  \caption{Comparison of ground-truth MD results with DDD predictions of [001] tension and compression deformation of BCC W using the GNN-based mobility trained from large-scale MD data. MD and DDD simulation volumes of identical size are deformed under identical uniaxial straining rates of $2\times 10^8$/s. (a) Axial stress and (b) dislocation density evolution as a function of strain.}
\label{fig:ddd_md_results}
\end{figure}

Results of the DDD predictions compared to MD simulations are shown in Fig.~\ref{fig:ddd_md_results}. Flow stress predictions are in remarkable agreement with the MD simulations, Fig.~\ref{fig:ddd_md_results}a. Specifically, the asymmetry in the tension/compression response is naturally captured by the GNN mobility law, demonstrating that the model has correctly learned the distinct behavior of the dislocations depending on shearing direction. 
In Fig.~\ref{fig:ddd_md_results}b however, we note a quantitative difference in the predicted evolution of the dislocation densities. While DDD indeed predicts a higher dislocation density in compression compared to tension, in agreement with the MD data, DDD also systematically underpredicts the value of the dislocation density by a factor of approximately two.
Potential causes of these discrepancies will be discussed in the next section.

\section{Discussion}

Overall, results presented in \S\ref{sec:ddd_md_results} clearly demonstrate the feasibility and potential of the approach.
These results are all the more significant that our previous attempts at developing traditional ``hand-crafted'' mobility laws to satisfactorily capture the tension/compression asymmetry in BCC metals have proved surprisingly challenging, with mixed results when comparing large-scale DDD predictions to reference MD simulations.
Indeed, the strong anisotropic plasticity of BCC metals, rooted in the T/AT asymmetry \cite{duesbery1998plastic, ito2001atomistic, vitek2004core, dezerald2016plastic}, translates into a complex motion of the dislocations as a function of the local configuration and stress state.  
Here, we show that such a behavior can be naturally captured by our ML approach. Through seeing a large number of examples during training, the GNN-based model learns to predict accurate, effective velocities, which in our example generally result in smooth stress-velocity relations, e.g. see Fig.~\ref{fig:md_mob}.

Of particular interest, the GNN mobility law also encodes information about slip crystallography (i.e. direction of the velocity vector), without pre-existing assumption about slip planes, which is an unsettled and not fully understood issue in BCC metals \cite{weinberger2013slip}. This is in contrast with earlier DDD mobility law frameworks that assume a fixed set of glide planes and only focus on modeling the magnitude of the dislocation velocity vector \cite{wang2011atomistically, po2016phenomenological}.
This example illustrates the novelty and power of such a tool to encode complex dislocation physics into the DDD model, without imposing constraints (e.g. functional forms, slip planes, etc.) that may prove detrimental.

In our first application, we nevertheless observe discrepancies compared to the ground-truth MD simulations.
While the flow stress in compression is remarkably captured within statistical fluctuations, flow stress under tension is slightly underestimated. Yet, as noted previously the largest discrepancy is found in the evolution of the dislocation densities, which are roughly underestimated by a factor two in both DDD simulations.

Several reasons could be at the origin of these discrepancies. The first set of reasons could be related to the DDD model itself. In DDD, dislocation mobility is not the only ingredient by which the dislocation network is evolved. Indeed, topological operations performed to handle core reactions (e.g. junction formation and splitting) constitute another critical step of the DDD method \cite{Arsenlis07}. This seperate step is performed in turn with the integration of the equation of motion, and core reactions are thus absent from the learning procedure. It is thus possible that the lack of multiplication observed in Fig.~\ref{fig:ddd_md_results}b is rather a consequence of the details of the treatment of topological operations than due to inaccuracies of the mobility law itself. An example of this issue was recently uncovered in \cite{bertin2022enhanced}.

The second reason could be a more general one, analogous to the issue of uncertainty quantification and transferability of interatomic potentials \cite{chernatynskiy2013uncertainty}.
For instance, it is possible that -- if it were feasible -- large-scale \emph{ab-initio} simulations would produce ground-truth outcomes that differ significantly from MD predictions using even the most accurate interatomic potentials fitted to it. In other words, if the emerging behavior (e.g. crystal plasticity) is sensitive to the smallest details (e.g. details of the electronic structure), then the transferabilty of physics across the scales may be a daunting task. By analogy a similar difficulty may exist in coarse-graining lattice motion to a dislocation mobility law.

As a way to address this issue, recent work has focused on the selection of the training data \cite{jeong2018toward, imbalzano2018automatic, podryabinkin2019accelerating, bernstein2019novo, montes2022training}, which has been an everlasting challenge in the realm of interatomic potentials. As a first application of our DDD+ML framework, here we have used training data produced on a very narrow set of loading conditions (unique high straining rate in tension and compression).
\added{It is thus possible that a broader training dataset could help reduce the discrepancies. Simultaneously,} it is also possible that the applicability of our trained mobility may be limited to a narrow range of conditions as well.

\begin{figure}[t]
  \begin{center}
    \includegraphics[width=0.49\textwidth]{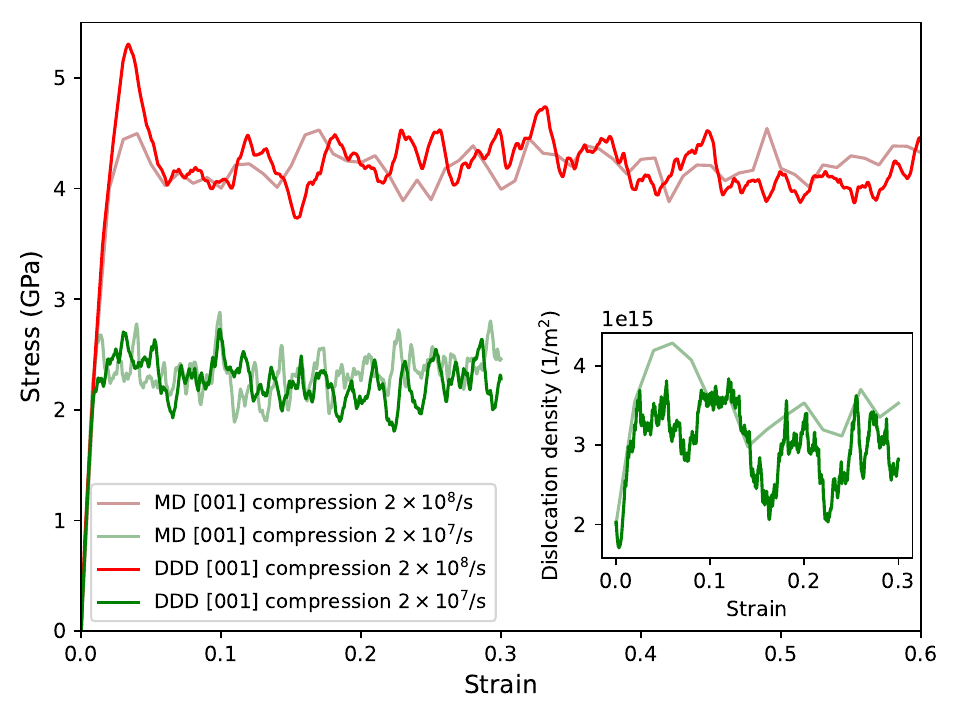}
  \end{center}
  \vspace*{-0.5cm}
  \caption{\added{Comparison of ground-truth MD results with DDD predictions of [001] compression for straining rates of $2\times 10^8$/s and $2\times 10^7$/s. Although training was only performed on $2\times 10^8$/s trajectories (red), our DDD+ML method also correctly predicts the flow stress at the lower straining rate of $2\times 10^7$/s (green). Comparison of the dislocation density is shown in the inset for the straining rate of $2\times 10^7$/s.}}
\label{fig:ddd_md_results_2e7}
\end{figure}

\added{To first examine the latter issue, we have run and compared results of simulations performed at a lower straining rate of $2\times 10^7$/s under compression. To ensure that volumes remain statically representative of bulk plasticity, the MD crystal size was increased by a factor 4 to $\sim 140$ millions atoms, i.e. corresponding to a DDD box of side length $470b$.
Results for the stress/strain predictions are shown in Fig.~\ref{fig:ddd_md_results_2e7}. Although the mobility law was only trained on $2\times 10^8$/s trajectories, the DDD+ML model is seen to remarkably capture the flow stress at this lower strain rate.
This result is consistent with our previous attempt at employing GNN to evolve simplified DDD systems \cite{bertin2023accelerating}, in which we similarly observed that the GNN model trained on high flow stress (i.e. high strain rate) simulations was able to correctly predict the evolution of simulations at lower strain rates not seen during training.
We believe one reason for this success could be that, despite the system being driven at a high external stress in the training data,} local dislocation lines sample a much richer set of stress conditions. Indeed, the stress acting along dislocations lines is not just the applied stress, but the sum of the applied stress and the internal stress arising from interactions with all other dislocations (e.g. short-range interactions).
While we leave further aspects of this issue out the scope of this paper, future work will focus on the accuracy and generalizability of the approach to different conditions.

\added{Finally, we point out that, as coarse-grained model, DDD simulations are typically several orders of magnitude faster than MD simulations under identical conditions. In DDD+ML, the GNN-based mobility only comes with a marginal cost overhead compared to traditional mobility laws, and the simulation cost is still fully dominated by force calculations. For large-scale examples presented in this work, MD simulations (at $2\times 10^8$/s) typically require $\sim 100-1000$ GPU-hours (e.g. depending on the interatomic potential), compared to $\sim 10-20$ CPU-hours needed for the same DDD simulations. At lower straining rates the gains are only even larger, and DDD can access volume sizes (e.g. several $\mu$m) that are out-of-reach from direct MD.}

To summarize, we have introduced a ML framework to construct DDD mobility laws trained on large-scale MD data using a GNN model. Training is achieved by matching the evolution of the continuum dislocation density tensor between subsequent DXA configurations for which an estimate of the dislocation driving forces computed with DDD is provided as input. This scheme allows us to circumvent both the issues of unknown nodal force and velocity vectors in the ground-truth MD data.
By applying our approach to the complex case of BCC W, we showed that \added{DDD+ML} simulations using a trained GNN mobility are able to reproduce well the behavior of ground-truth MD simulations it was trained on, \added{and that of unseen loading conditions}.
We believe that our proposed approach to construct data-driven mobility laws is a promising avenue that has the potential to significantly improve the fidelity of the DDD model while allowing to incorporate more complex dislocation motion behaviors which, if desired, can be made unbiased of human intervention.

\section{Methods}

\subsection{GNN mobility law}

We model the mobility law in Eq.~\eqref{eq:M_GNN} with a message-passing GNN \cite{battaglia2018relational}, which has proved very powerful for predicting force fields and other materials properties \cite{gilmer2017neural, Xie2018, Chen2018, Park2021, Batzner2021} and for simulating complex physics \cite{sanchez2020learning}. Following our recent work on applying GNN to DDD simulations \cite{bertin2023accelerating}, a dislocation configuration is represented by a graph $\mathcal{G} = (\mathcal{V},\mathcal{E})$, where $\mathcal{V} = \{N_i\}$ is a collection of dislocation node/vertex features (attributes), and $\mathcal{E} = \{E_{ij}\}$ is a collection of dislocation segment/edge features.
We define the input features for each node $i$ as
\begin{equation} \label{eq:node-feature}
N_i = (n_i,  \vect{F}^{\rm DDD}, \|\vect{F}^{\rm DDD}\| )
\end{equation}
where $n_i$ is a flag used to specify whether node $i$ is a discretization or physical (junction) node.
The input edge features on edge $ij$ are
\begin{equation} \label{eq:edge-feature}
E_{ij} =( e_{ij}, \vect{b}_{ij}, \|\vect{b}_{ij}\|, \vect{r}_{j}-\vect{r}_{i}, \|\vect{r}_{j}-\vect{r}_{i}\| )
\end{equation}
where $e_{ij}$ is a flag used to specify whether segment $ij$ is a glissile or junction segment, and $\vect{r}_{j}-\vect{r}_{i}$ are the local segments line vectors, naturally compatible with the use of periodic boundary conditions.
To satisfy Burgers vector conservation, we use a directed graph, i.e. if $ij$ is an edge with Burgers vector $\vect{b}_{ij}$, then $ji$ is also an edge but with opposite Burgers vector $-\vect{b}_{ij}$.

Our GNN architecture follows \cite{MeshGraphNet} and is first composed of vertex and edge encoders $\text{ENC}^V$, $\text{ENC}^E$ transforming concatenated input features into a latent space
\begin{equation} \label{eq:encoder}
    v^{(0)}_i = \text{ENC}^V(N_i), \ e^{(0)}_{ij} = \text{ENC}^E(E_{ij}),
\end{equation}
followed by $K$ stacked message passing layers $f^{E(k)}$, $f^{V(k)}$ ($1 \leq k \leq K$) sequentially updating the latent vertex and edge variables
\begin{align}
e^{(k)}_{ij} &= f^{E(k)}( e^{(k-1)}_{ij}, v^{(k-1)}_i, v^{(k-1)}_j ),  \\
v^{(k)}_{i} &= f^{V(k)}( v^{(k-1)}_i, \sum_j e^{(k)}_{ij}), \label{eq:MP}
\end{align}
and finally a node decoder $\text{DEC}$ that translates the latent node variables $v^{(K)}$ into the desired output properties, i.e. nodal velocity vectors:
\begin{equation} \label{eq:decoder}
    \vect{V}_i  = \text{DEC}(v^{(K)}_i).
\end{equation}
Functions $\text{ENC}^V$, $\text{ENC}^E$, $f^V$, $f^E$, and $\text{DEC}$ are neural network operators built from multi-layer perceptrons with two hidden layers, layer normalization \cite{LayerNorm}, skip connections \cite{resnet}, and GELU activation functions \cite{GELU}.

\subsection{MD simulations}

Large-scale MD simulations of BCC tungsten are performed with LAMMPS \cite{thompson2022lammps} under the Kokkos GPU implementation \cite{edwards2014kokkos}, using the EAM-style interatomic potential developed in \cite{Juslin13}. Simulations are performed following the protocol introduced in \cite{zepeda2017probing}. Periodic, orthorombic BCC perfect crystals are initially seeded with twelve $1/2\langle 111 \rangle \{110\}$ prismatic loops of the vacancy type.
The crystals are first equilibrated at the temperature of 300~K, after which they are deformed at a constant true strain rate.
Temperature and uniaxial loading conditions are maintained during deformation using the \emph{langevin} thermostat and the \emph{nph} barostat. \added{For the $\sim 35$ millions atoms simulations deformed at a rate of $2 \times 10^8$/s}, DXA \cite{Stukowski14} is executed every 1~ps to save the detailed evolution of the dislocation networks.

\subsection{Mobility workflow and training}

DXA configurations produced in the MD simulations are first converted to the ParaDiS format \cite{Arsenlis07}. For consistency, during this operation the line networks are also remeshed with discretization size of $\sim 10b$, corresponding to the average segment size in used our DDD simulations.

The so-converted dislocation configurations $\{\mathcal{G}^{t_s}\}$ are then fed to the ParaDiS code to compute nodal forces, Eq.~\eqref{eq:Fi_DDD}.
Applied forces $\vect{F}_i^{app}$ are computed by integrating the Peach-Koehler force $(\tensor{\sigma}^{t_s} \cdot \vect{b}_{ij}) \times \vect{t}_{ij}$ along the dislocation segments $ij$ with unit tangent $\vect{t}_{ij}$, where $\tensor{\sigma}^{t_s}$ is the instantaneous stress applied to network $\mathcal{G}^{t_s}$ at time $t_s$ as recorded during the MD runs.
Long-range interaction forces $\vect{F}_i^{lr}$ are computed using DDD-FFT approach introduced in \cite{bertin2019connecting} which can easily handle non-cubic, deforming simulation boxes as produced by MD simulations.
Short-range interaction forces $\vect{F}_i^{sr}$ are computed using the non-singular isotropic analytical formulation \cite{cai2006non}.
Core forces $\vect{F}_i^{core}$ are computed from core energies extracted from the ground-truth interatomic potential \cite{Juslin13} using the framework developed in \cite{bertin2021core}.

The networks containing nodal forces are then used as inputs to our training procedure implemented within PyTorch \cite{pytorch}. To facilitate training, input forces are rescaled by a factor $10^9~{\rm Pa}\cdot b^2$ so that their average magnitude is on the order of unity. The loss function, Eq.~\eqref{eq:loss}, is computed by evaluating the Nye's tensor on a grid of $32^3$ voxels using a fully vectorized implementation of the discrete-to-continuous method introduced in \cite{bertin2019connecting}.

We trained different GNN models to explore different sets of hyper-parameters. We tested a combination of models with $K=\{2,3\}$ message-passing layers each with latent space of size $L=\{48,96\}$, leading to 4 different trained models. The models were trained using the Adamw optimizer with weight decay of $1 \times 10^{-5}$ \cite{Adamw}. Training was performed for 12 hours with batch size of 4 on a single NVidia V100 GPU.
We find that the GNN model with $K=3$ and $L=48$ offers the best trade-off between accuracy and complexity (78,768 total parameters) while showing no sign of ovefitting. We thus selected it as the best model for results presented in this work.

\section*{Acknowledgement}
NB and VB acknowledge support by the Laboratory Directed Research and Development (LDRD) program (22-ERD-016) and by the ASC PEM program at Lawrence Livermore National Laboratory (LLNL). FZ was supported by the Critical Materials Institute, an Energy Innovation Hub funded by the U.S. Department of Energy, Office of Energy Efficiency and Renewable Energy, and Advanced Materials and Manufacturing Technologies Office.
Computing support for this work came from LLNL Institutional Computing Grand Challenge program.
This work was performed under the auspices of the U.S. Department of Energy by LLNL under contract DE-AC52-07NA27344.

\appendix


\bibliography{ref}

\begin{thebibliography}{10}
\expandafter\ifx\csname url\endcsname\relax
  \def\url#1{\texttt{#1}}\fi
\expandafter\ifx\csname urlprefix\endcsname\relax\def\urlprefix{URL }\fi
\expandafter\ifx\csname href\endcsname\relax
  \def\href#1#2{#2} \def\path#1{#1}\fi

\bibitem{zepeda2017probing}
L.~A. Zepeda-Ruiz, A.~Stukowski, T.~Oppelstrup, V.~V. Bulatov, Probing the
  limits of metal plasticity with molecular dynamics simulations, Nature
  550~(7677) (2017) 492--495.

\bibitem{zepeda2021atomistic}
L.~A. Zepeda-Ruiz, A.~Stukowski, T.~Oppelstrup, N.~Bertin, N.~R. Barton,
  R.~Freitas, V.~V. Bulatov, Atomistic insights into metal hardening, Nature
  materials 20~(3) (2021) 315--320.

\bibitem{bertin2022sweep}
N.~Bertin, L.~Zepeda-Ruiz, V.~Bulatov, Sweep-tracing algorithm: in silico slip
  crystallography and tension-compression asymmetry in bcc metals, Materials
  Theory 6~(1) (2022) 1--23.

\bibitem{stimac2022energy}
J.~C. Stimac, N.~Bertin, J.~K. Mason, V.~V. Bulatov, Energy storage under
  high-rate compression of single crystal tantalum, Acta Materialia 239 (2022)
  118253.

\bibitem{kubin1992dislocation}
L.~P. Kubin, G.~Canova, M.~Condat, B.~Devincre, V.~Pontikis, Y.~Br{\'e}chet,
  Dislocation microstructures and plastic flow: a 3d simulation, Solid state
  phenomena 23 (1992) 455--472.

\bibitem{zbib1998plastic}
H.~M. Zbib, M.~Rhee, J.~P. Hirth, On plastic deformation and the dynamics of 3d
  dislocations, International Journal of Mechanical Sciences 40~(2-3) (1998)
  113--127.

\bibitem{bulatov1998connecting}
V.~Bulatov, F.~F. Abraham, L.~Kubin, B.~Devincre, S.~Yip, Connecting atomistic
  and mesoscale simulations of crystal plasticity, Nature 391~(6668) (1998)
  669--672.

\bibitem{schwarz1999simulation}
K.~Schwarz, Simulation of dislocations on the mesoscopic scale. i. methods and
  examples, Journal of Applied Physics 85~(1) (1999) 108--119.

\bibitem{ghoniem2000parametric}
N.~Ghoniem, M, S.-H. Tong, L.~Sun, Parametric dislocation dynamics: a
  thermodynamics-based approach to investigations of mesoscopic plastic
  deformation, Physical Review B 61~(2) (2000) 913.

\bibitem{weygand2002aspects}
D.~Weygand, L.~Friedman, E.~Van~der Giessen, A.~Needleman, Aspects of
  boundary-value problem solutions with three-dimensional dislocation dynamics,
  Modelling and Simulation in Materials Science and Engineering 10~(4) (2002)
  437.

\bibitem{Arsenlis07}
A.~Arsenlis, W.~Cai, M.~Tang, M.~Rhee, T.~Oppelstrup, G.~Hommes, T.~G. Pierce,
  V.~V. Bulatov, Enabling strain hardening simulations with dislocation
  dynamics, Modelling and Simulation in Materials Science and Engineering
  15~(6) (2007) 553--595.

\bibitem{BulatovCai06}
V.~Bulatov, W.~Cai, Computer simulations of dislocations, Vol.~3, Oxford
  University Press on Demand, 2006.

\bibitem{wang2011atomistically}
Z.~Wang, I.~Beyerlein, An atomistically-informed dislocation dynamics model for
  the plastic anisotropy and tension--compression asymmetry of bcc metals,
  International Journal of Plasticity 27~(10) (2011) 1471--1484.

\bibitem{srivastava2013dislocation}
K.~Srivastava, R.~Gr{\"o}ger, D.~Weygand, P.~Gumbsch, Dislocation motion in
  tungsten: atomistic input to discrete dislocation simulations, International
  Journal of Plasticity 47 (2013) 126--142.

\bibitem{po2016phenomenological}
G.~Po, Y.~Cui, D.~Rivera, D.~Cereceda, T.~D. Swinburne, J.~Marian, N.~Ghoniem,
  A phenomenological dislocation mobility law for bcc metals, Acta Materialia
  119 (2016) 123--135.

\bibitem{chang2001dislocation}
J.~Chang, W.~Cai, V.~V. Bulatov, S.~Yip, Dislocation motion in bcc metals by
  molecular dynamics, Materials Science and Engineering: A 309 (2001) 160--163.

\bibitem{olmsted2005atomistic}
D.~L. Olmsted, L.~G. Hector, W.~Curtin, R.~Clifton, Atomistic simulations of
  dislocation mobility in al, ni and al/mg alloys, Modelling and Simulation in
  Materials Science and Engineering 13~(3) (2005) 371.

\bibitem{queyreau2011edge}
S.~Queyreau, J.~Marian, M.~Gilbert, B.~Wirth, Edge dislocation mobilities in
  bcc fe obtained by molecular dynamics, Physical Review B 84~(6) (2011)
  064106.

\bibitem{cereceda2013assessment}
D.~Cereceda, A.~Stukowski, M.~Gilbert, S.~Queyreau, L.~Ventelon, M.-C.
  Marinica, J.~Perlado, J.~Marian, Assessment of interatomic potentials for
  atomistic analysis of static and dynamic properties of screw dislocations in
  w, Journal of Physics: Condensed Matter 25~(8) (2013) 085702.

\bibitem{cho2017mobility}
J.~Cho, J.-F. Molinari, G.~Anciaux, Mobility law of dislocations with several
  character angles and temperatures in fcc aluminum, International Journal of
  Plasticity 90 (2017) 66--75.

\bibitem{monnet2009structure}
G.~Monnet, D.~Terentyev, Structure and mobility of the 12< 111>$\{$112$\}$ edge
  dislocation in bcc iron studied by molecular dynamics, Acta Materialia 57~(5)
  (2009) 1416--1426.

\bibitem{kang2012singular}
K.~Kang, V.~V. Bulatov, W.~Cai, Singular orientations and faceted motion of
  dislocations in body-centered cubic crystals, Proceedings of the National
  Academy of Sciences 109~(38) (2012) 15174--15178.

\bibitem{geslin2018thermal}
P.-A. Geslin, D.~Rodney, Thermal fluctuations of dislocations reveal the
  interplay between their core energy and long-range elasticity, Physical
  Review B 98~(17) (2018) 174115.

\bibitem{bertin2020frontiers}
N.~Bertin, R.~B. Sills, W.~Cai, Frontiers in the simulation of dislocations,
  Annual Review of Materials Research 50 (2020) 437--464.

\bibitem{sills2018dislocation}
R.~B. Sills, N.~Bertin, A.~Aghaei, W.~Cai, Dislocation networks and the
  microstructural origin of strain hardening, Physical review letters 121~(8)
  (2018) 085501.

\bibitem{bertin2023accelerating}
N.~Bertin, F.~Zhou, Accelerating discrete dislocation dynamics simulations with
  graph neural networks, Journal of Computational Physics 487 (2023) 112180.

\bibitem{Stukowski10}
A.~Stukowski, K.~Albe, Extracting dislocations and non-dislocation crystal
  defects from atomistic simulation data, Modelling and Simulation in Materials
  Science and Engineering 18~(8) (2010) 085001.

\bibitem{Stukowski14}
A.~Stukowski, A triangulation-based method to identify dislocations in
  atomistic models, Journal of the Mechanics and Physics of Solids 70 (2014)
  314--319.

\bibitem{Nye53}
J.~Nye, Some geometrical relations in dislocated crystals, Acta metallurgica
  1~(2) (1953) 153--162.

\bibitem{bertin2019connecting}
N.~Bertin, Connecting discrete and continuum dislocation mechanics: A
  non-singular spectral framework, International Journal of Plasticity 122
  (2019) 268--284.

\bibitem{pytorch}
A.~Paszke, S.~Gross, F.~Massa, A.~Lerer, J.~Bradbury, G.~Chanan, T.~Killeen,
  Z.~Lin, N.~Gimelshein, L.~Antiga, A.~Desmaison, A.~Kopf, E.~Yang, Z.~DeVito,
  M.~Raison, A.~Tejani, S.~Chilamkurthy, B.~Steiner, L.~Fang, J.~Bai,
  S.~Chintala, Pytorch: An imperative style, high-performance deep learning
  library, in: Advances in Neural Information Processing Systems 32, Curran
  Associates, Inc., 2019, pp. 8024--8035.

\bibitem{cai2004mobility}
W.~Cai, V.~V. Bulatov, Mobility laws in dislocation dynamics simulations,
  Materials Science and Engineering: A 387 (2004) 277--281.

\bibitem{christian1983some}
J.~Christian, Some surprising features of the plastic deformation of
  body-centered cubic metals and alloys, Metallurgical Transactions A 14~(7)
  (1983) 1237--1256.

\bibitem{duesbery1998plastic}
M.~a.-S. Duesbery, V.~Vitek, Plastic anisotropy in bcc transition metals, Acta
  Materialia 46~(5) (1998) 1481--1492.

\bibitem{dezerald2016plastic}
L.~Dezerald, D.~Rodney, E.~Clouet, L.~Ventelon, F.~Willaime, Plastic anisotropy
  and dislocation trajectory in bcc metals, Nature Communications 7 (2016)
  11695.

\bibitem{sherwood1967plastic}
P.~Sherwood, F.~Guiu, H.~C. Kim, P.~L. Pratt, Plastic anisotropy of tantalum,
  niobium, and molybdenum, Canadian Journal of Physics 45~(2) (1967)
  1075--1089.

\bibitem{webb1974effect}
G.~L. Webb, R.~Gibala, T.~E. Mitchell, Effect of normal stress on yield
  asymmetry in high purity tantalum crystals, Metallurgical Transactions 5~(7)
  (1974) 1581--1584.

\bibitem{weinberger2013slip}
C.~R. Weinberger, B.~L. Boyce, C.~C. Battaile, Slip planes in bcc transition
  metals, International Materials Reviews 58~(5) (2013) 296--314.

\bibitem{Juslin13}
N.~Juslin, B.~Wirth, Interatomic potentials for simulation of he bubble
  formation in w, Journal of nuclear materials 432~(1-3) (2013) 61--66.

\bibitem{bertin2023crystal}
N.~Bertin, R.~Carson, V.~V. Bulatov, J.~Lind, M.~Nelms, Crystal plasticity
  model of bcc metals from large-scale md simulations, Acta Materialia 260
  (2023) 119336.

\bibitem{osetsky2003atomic}
Y.~N. Osetsky, D.~J. Bacon, An atomic-level model for studying the dynamics of
  edge dislocations in metals, Modelling and simulation in materials science
  and engineering 11~(4) (2003) 427.

\bibitem{edagawa1997motion}
K.~Edagawa, T.~Suzuki, S.~Takeuchi, Motion of a screw dislocation in a
  two-dimensional peierls potential, Physical Review B 55~(10) (1997) 6180.

\bibitem{bertin2021core}
N.~Bertin, W.~Cai, S.~Aubry, V.~Bulatov, Core energies of dislocations in bcc
  metals, Physical Review Materials 5~(2) (2021) 025002.

\bibitem{ito2001atomistic}
K.~Ito, V.~Vitek, Atomistic study of non-schmid effects in the plastic yielding
  of bcc metals, Philosophical Magazine A 81~(5) (2001) 1387--1407.

\bibitem{vitek2004core}
V.~Vitek, Core structure of screw dislocations in body-centred cubic metals:
  relation to symmetry and interatomic bonding, Philosophical Magazine 84~(3-5)
  (2004) 415--428.

\bibitem{bertin2022enhanced}
N.~Bertin, W.~Cai, S.~Aubry, A.~Arsenlis, V.~V. Bulatov, Enhanced mobility of
  dislocation network nodes and its effect on dislocation multiplication and
  strain hardening, arXiv preprint arXiv:2210.14343 (2022).

\bibitem{chernatynskiy2013uncertainty}
A.~Chernatynskiy, S.~R. Phillpot, R.~LeSar, Uncertainty quantification in
  multiscale simulation of materials: A prospective, Annual Review of Materials
  Research 43 (2013) 157--182.

\bibitem{jeong2018toward}
W.~Jeong, K.~Lee, D.~Yoo, D.~Lee, S.~Han, Toward reliable and transferable
  machine learning potentials: uniform training by overcoming sampling bias,
  The Journal of Physical Chemistry C 122~(39) (2018) 22790--22795.

\bibitem{imbalzano2018automatic}
G.~Imbalzano, A.~Anelli, D.~Giofr{\'e}, S.~Klees, J.~Behler, M.~Ceriotti,
  Automatic selection of atomic fingerprints and reference configurations for
  machine-learning potentials, The Journal of chemical physics 148~(24) (2018).

\bibitem{podryabinkin2019accelerating}
E.~V. Podryabinkin, E.~V. Tikhonov, A.~V. Shapeev, A.~R. Oganov, Accelerating
  crystal structure prediction by machine-learning interatomic potentials with
  active learning, Physical Review B 99~(6) (2019) 064114.

\bibitem{bernstein2019novo}
N.~Bernstein, G.~Cs{\'a}nyi, V.~L. Deringer, De novo exploration and
  self-guided learning of potential-energy surfaces, npj Computational
  Materials 5~(1) (2019) 99.

\bibitem{montes2022training}
D.~Montes~de Oca~Zapiain, M.~A. Wood, N.~Lubbers, C.~Z. Pereyra, A.~P.
  Thompson, D.~Perez, Training data selection for accuracy and transferability
  of interatomic potentials, npj Computational Materials 8~(1) (2022) 189.

\bibitem{battaglia2018relational}
P.~W. Battaglia, J.~B. Hamrick, V.~Bapst, A.~Sanchez-Gonzalez, V.~Zambaldi,
  M.~Malinowski, A.~Tacchetti, D.~Raposo, A.~Santoro, R.~Faulkner, et~al.,
  Relational inductive biases, deep learning, and graph networks (2018).
\newblock \href {http://arxiv.org/abs/1806.01261} {\path{arXiv:1806.01261}}.

\bibitem{gilmer2017neural}
J.~Gilmer, S.~S. Schoenholz, P.~F. Riley, O.~Vinyals, G.~E. Dahl, Neural
  message passing for quantum chemistry, in: International conference on
  machine learning, PMLR, 2017, pp. 1263--1272.

\bibitem{Xie2018}
T.~Xie, J.~C. Grossman, {Crystal Graph Convolutional Neural Networks for an
  Accurate and Interpretable Prediction of Material Properties}, Physical
  Review Letters 120~(14) (2018) 145301.

\bibitem{Chen2018}
C.~Chen, W.~Ye, Y.~Zuo, C.~Zheng, S.~P. Ong, {Graph Networks as a Universal
  Machine Learning Framework for Molecules and Crystals}, Chemistry of
  Materials 31 (2018) 3564.

\bibitem{Park2021}
C.~W. Park, M.~Kornbluth, J.~Vandermause, C.~Wolverton, B.~Kozinsky, J.~P.
  Mailoa, {Accurate and scalable graph neural network force field and molecular
  dynamics with direct force architecture}, npj Computational Materials 7~(1)
  (2021) 73.

\bibitem{Batzner2021}
S.~Batzner, A.~Musaelian, L.~Sun, M.~Geiger, J.~P. Mailoa, M.~Kornbluth,
  N.~Molinari, T.~E. Smidt, B.~Kozinsky, {E(3)-equivariant graph neural
  networks for data-efficient and accurate interatomic potentials}, Nature
  Communications 13~(1) (2022) 2453.

\bibitem{sanchez2020learning}
A.~Sanchez-Gonzalez, J.~Godwin, T.~Pfaff, R.~Ying, J.~Leskovec, P.~Battaglia,
  Learning to simulate complex physics with graph networks, in: International
  Conference on Machine Learning, PMLR, 2020, pp. 8459--8468.

\bibitem{MeshGraphNet}
T.~Pfaff, M.~Fortunato, A.~Sanchez-Gonzalez, P.~W. Battaglia, {Learning
  mesh-based simulation with graph networks} (2020).
\newblock \href {http://arxiv.org/abs/2010.03409} {\path{arXiv:2010.03409}}.

\bibitem{LayerNorm}
J.~L. Ba, J.~R. Kiros, G.~E. Hinton, Layer normalization (2016).
\newblock \href {http://arxiv.org/abs/1607.06450} {\path{arXiv:1607.06450}}.

\bibitem{resnet}
K.~He, X.~Zhang, S.~Ren, J.~Sun, Deep residual learning for image recognition
  (2015).
\newblock \href {http://arxiv.org/abs/1512.03385} {\path{arXiv:1512.03385}}.

\bibitem{GELU}
D.~Hendrycks, K.~Gimpel, Gaussian error linear units (gelus) (2016).
\newblock \href {http://arxiv.org/abs/1606.08415} {\path{arXiv:1606.08415}}.

\bibitem{thompson2022lammps}
A.~P. Thompson, H.~M. Aktulga, R.~Berger, D.~S. Bolintineanu, W.~M. Brown,
  P.~S. Crozier, P.~J. in't Veld, A.~Kohlmeyer, S.~G. Moore, T.~D. Nguyen,
  et~al., Lammps-a flexible simulation tool for particle-based materials
  modeling at the atomic, meso, and continuum scales, Computer Physics
  Communications 271 (2022) 108171.

\bibitem{edwards2014kokkos}
H.~C. Edwards, C.~R. Trott, D.~Sunderland, Kokkos: Enabling manycore
  performance portability through polymorphic memory access patterns, Journal
  of parallel and distributed computing 74~(12) (2014) 3202--3216.

\bibitem{cai2006non}
W.~Cai, A.~Arsenlis, C.~R. Weinberger, V.~V. Bulatov, A non-singular continuum
  theory of dislocations, Journal of the Mechanics and Physics of Solids 54~(3)
  (2006) 561--587.

\bibitem{Adamw}
I.~Loshchilov, F.~Hutter, Decoupled weight decay regularization (2017).
\newblock \href {http://arxiv.org/abs/1711.05101} {\path{arXiv:1711.05101}}.

\end{thebibliography}

\end{document}